\documentstyle[aps,prl,epsf,epsfig,floats]{revtex}
\begin{document}
\draft

\title{Current oscillations in a metallic ring
threaded by a time-dependent magnetic flux.}
\author{Liliana Arrachea}
\address{Scuola Internazionale Superiore di Studi Avanzati\\
Via Beirut n. 2-4 (34013) Trieste, Italy.\\
 and \\
Departamento de F\'{\i}sica FCEyN, 
Universidad de Buenos Aires\\
Pabell\'{o}n I, Ciudad Universitaria (1428) Buenos Aires, 
Argentina.}

\maketitle

\begin{abstract}
We study a mesoscopic metallic ring threaded by a magnetic
flux which varies linearly in time $\Phi_M(t)=\Phi t$
with a formalism based in Baym-Kadanoff-Keldysh non-equilibrium 
Green functions.
We propose a method to calculate the Green functions in real space and 
we consider an experimental setup to investigate the dynamics of the
ring by recourse to a transport experiment. This consists in
a single lead connecting the ring to a particle reservoir. 
We show that different
dynamical regimes are attained depending on the ratio 
$\hbar \Phi/\Phi_0 W$, being $\Phi_0=h c/e$ and 
$W$, the bandwidth of the ring.
For moderate lengths of the ring,   
a stationary regime is achieved for   $\hbar \Phi/\Phi_0 >W$. 
In the opposite case with $\hbar \Phi/\Phi_0 \leq W$, 
the effect of Bloch oscillations driven by the 
induced electric field manifests itself in the transport 
properties of the system. In particular, we show that in
this time-dependent regime a tunneling current oscillating in time 
with  a period $\tau=2\pi\Phi_0/\Phi$ can be measured in the lead.
We also analyze the resistive effect introduced by inelastic 
scattering due to the coupling to the external reservoir. 
\end{abstract}


\section{Introduction}
In the last years, the progress of technology
has enabled the experimental study of very  
interesting phenomena taking place in mesoscopic systems. 
The fascinating feature in these systems is that
several physical processes that have a 
pure quantum-mechanical origin can be observed 
in a suitable transport experiment. 
Paradigmatic examples are the 
Aharanov-Bohm effect in systems with annular geometry
and Bloch oscillations in periodic structures.

The Aharanov-Bohm effect in circular devices  
 motivated a vast  theoretical and experimental activity 
for some time now.
A metallic mesoscopic ring threaded by a static magnetic
field displays persistent currents, which oscillate 
as a function of the magnetic flux $\Phi_M$, with a 
period $\Phi_0=2 \pi h/e$ \cite{percur}.
Like other mesoscopic phenomena, 
these persistent currents are a consequence of the coherence
of the wave function throughout the system.
The Aharanov-Bohm effect also manifests itself when such a
ring is connected to two metallic leads at different external 
voltages. This geometry is usually employed in experiments
to measure the conductance of the metallic ring in the presence of the
external flux. In this case, quantum interference cause oscillations in the
current through the collector
 lead  with a period $\Phi_0$ as $\Phi_M$
varies \cite{intab}. 

More than 70 years ago, Bloch predicted the occurrence of a
time-periodic motion of the electrons in a band under the influence
of an external constant electric field $F$ \cite{blo}.
Recently,  Bloch oscillations with a period $h/eFL$ has been observed 
in
 superlattices \cite{bloex},
being $L$ the periodicity of the lattice and $e$ the electron 
charge. This motivated
new theoretical development in the field
\cite{blot1,blot2,blot3}.

A metallic ring penetrated by a time-dependent magnetic field is a
very interesting system, 
worth of theoretical and experimental study. 
Since time is the
unavoidable variable in any experimental measurement
of currents while the magnetic field is varied, that device is more
realistic than its static-field counterpart.
The case of a magnetic flux with a linear time dependence 
is particularly appealing because this situation is
closely related with that of electrons in a periodic lattice
under the influence of a dc voltage. In fact,
a magnetic flux that varies linearly in time induces
a constant electric field $F=(1/cL) d\Phi_M(t)/dt$, being $L$ 
the longitude of the
 ring, while the geometry of the ring produces a periodic
structure for the electrons inside it.  
While a good amount of literature can be found on the Aharanov-Bohm effect,
considerably fewer studies have been reported on this problem. It
has been addressed in an early  work by B\"uttiker, Imry and Landauer
\cite{bil} who predicted that the electrons in an ideal one-dimensional ring
should execute Bloch  oscillations driven by the induced electric field. Some
time later,  Landauer and B\"uttiker showed that inelastic scattering in that
system generates a dc current and dissipation \cite{lb}. A single lead
connected between the ring and an external reservoir was later proposed
 by  B\"uttiker as a concrete way to introduce inelastic scattering events
\cite{b}. Since then, the attention focused  mainly in Aharanov-Bohm loops
where the flux threading the ring is static and devices 
with inserted quantum dots seem to occupy the center of the stage
\cite{yac}.   

In this work the problem of the normal-metal
 loop in the presence of an external 
time dependent magnetic flux with the form $\Phi_M(t)=\Phi t$ is studied in the
framework of non-equilibrium Green functions. We consider an ideal
ring described by a  one-dimensional tight-binding model with a bandwidth
$W$ and we propose an approach to calculate
the Green functions in real
space. This procedure  enables to gain insight on the different energy scales
that intervene in the  dynamics of the problem. In addition, it has the
advantage of being also adequate to tackle more general situations where
the configuration is not translational invariant, like the case of rings 
with disorder and with inserted barriers or quantum dots.

The experimental study of  the physical properties of the ring
in the presence of the variable magnetic field requires some kind of 
coupling between this system and the external world. 
We consider a concrete experimental setup to study time-dependent charge
oscillations in the ring. This consists in a single lead connecting it
to a particle reservoir.
The goal is to study the quantum tunneling between the ring and the 
reservoir, originated by the time dependent oscillations driven by 
the induced electric field.
The metallic wire is described as an ideal system of
 non-interacting electrons and
the reservoir is assumed to fix the chemical potential $\mu$. We use an
approach based in Baym-Kadanoff-Keldysh formalism to solve the coupled system.
This method has been successful to derive generalized transport
equations \cite{past,jau} in systems with an explicit time dependence. In the
present context the main idea is to represent the effect of the lead and the
reservoir by the self-energy effects they produce.

In the case of pure one-dimensional rings pierced by a
static magnetic field in the absence of disorder, the persistent 
currents vanish as $1/L$, 
being $L$ the circumference of the ring
\cite{ge}. In the time dependent case there is a non trivial connection between
the size of the system and its dynamical behavior.  We  
study the behavior of the retarded Green function
as a function of $\Phi$ and $L$. We argue that
for a magnetic field varying at a rate such that $ h \Phi/\Phi_0 \leq W$, 
the time dependent features of the relevant physical magnitudes 
 vanish as $1/L$. Instead, for more
rapid changes of the magnetic field such that $ h \Phi/\Phi_0 > W$,
their time dependence vanish exponentially fast as a function of $L$.
 Our aim is to show that for moderate lengths $L$
and field variations satisfying the condition $ h \Phi/\Phi_0 \leq W$,
quantum tunneling takes place between the ring and the
reservoir and a time dependent 
current could be measured in the lead.
We also analyze the effect that inelastic scattering originated by the 
coupling to the reservoir  cause on the transport properties of this system.

The paper is organized as follows. Section II is devoted to 
expose the theoretical approach. In Section III we obtain the set 
of equations to calculate the retarded Green function in the ring
unconnected from the reservoir as well as  connected to it.
Results are shown in Section IV. We finally summarize our
study and present the conclusions in Section V.  

\section{Basic theory}
We consider an experimental setup 
to study the  dynamics of the ring in the presence of the time-dependent
magnetic flux of the form $\Phi_M=\Phi t$. The device is sketched in Fig. 1a. 
Our aim is to study the quantum 
tunneling between the ring and an external particle reservoir by
measuring  the current flowing through the lead. 
We use an approach based in Baym-Kadanoff-Keldysh 
non-equilibrium Green functions to
solve the coupled system. We follow a procedure similar to that used in
Refs.\cite{past,jau}, where the effect of the lead and the reservoir is
represented by a suitable self-energy. The system 
is simplified 
as indicated in Fig. 1b. The
lead and the reservoir are described as an effective system $\alpha$ with a
fixed chemical potential $\mu$. A contact, described by a single bond,
 establishes the connection between the ring and the lead.

The total hamiltonian of this system contains three pieces:
\begin{equation}
H= H_{ring}+H_{\alpha}+H_{contact}.
\label{e18}
\end{equation}
The first term describes the metal ring threaded by the magnetic flux.
We consider  non-interacting spinless electrons
in  a tight-binding chain of longitude $L$ with $N$ sites, lattice parameter $a=L/N$
and  periodic  boundary conditions ($N+1 \equiv 1$).
The model hamiltonian is
\begin{equation}
H_{ring} \;=\; -T \sum_{l}^{N}  
( e^{-i \phi t} c^{\dagger}_l c_{l+1} +   
e^{i \phi t} c^{\dagger}_{l+1} c_l), 
\label{e1} 
\end{equation} 
with $l=1,\ldots,N$ and $\phi=\Phi/(\Phi_0 N)$, being $T$ the hopping matrix
element between nearest neighbors. The time-dependent phase $\phi t$ attached
to each link accounts for the  presence of the external magnetic flux.
The term $H_{\alpha}$ represents the lead and
the reservoir.
For the moment we do not give any explicit hamiltonian for this system.
 It is enough to assume that it is an infinite 
system of non-interacting electrons with a chemical potential $\mu$.
The hamiltonian describing the contact is
\begin{equation}
H_{contact}=-T_{1\alpha} 
(c^{\dagger}_1 c_{\alpha} +c^{\dagger}_{\alpha} c_1 )
\label{e19}.
\end{equation}

In terms of Green functions, the current through the contact is 
written as
\begin{equation}
J_{1\alpha}=\frac{2e}{\hbar} Re[ T_{1\alpha} G^{<}_{1\alpha}(t,t)],
\label{e20}
\end{equation}
where $G^<_{1 \alpha} (t, t')=i \langle c^{\dagger}_1 (t)  c_{ \alpha} (t') \rangle$.  
The latter obeys the following equation,
\begin{equation}
 G^<_{1 \alpha} (t, t')= -T_{1 \alpha}
\int dt_1 [G^R_{1,1}(t, t_1) g^<_{\alpha} (t_1- t^{\prime}) 
+ G^<_{1,1}(t, t_1) g^A_{\alpha} (t_1- t^{\prime}) ].
\label{e21}
\end{equation}
The functions 
$g^R_{\alpha} (t_1- t^{\prime}), g^A_{\alpha} (t_1-t^{\prime})$ 
and $g^<_{\alpha} (t_1- t^{\prime})$, with 
$g^{A}_{\alpha}(t-t^{\prime})=[g^R_{\alpha}(t^{\prime}-t)]^*$ are 
equilibrium Green functions corresponding to the noninteracting system
$\alpha$ before the coupling to the ring is performed.
In terms of the  spectral density 
$\rho_{\alpha} (\omega)= -2 Im[g^R_{\alpha}(\omega)]$ and the Fermi
function $f_{\alpha}(\omega)=\Theta(\mu-\omega)$ of this system,
they read 
\begin{eqnarray}
g^R_{\alpha}(t-t^{\prime}) &=& -i \Theta (t-t^{\prime}) \int 
\frac{d \omega}{2\pi}
  \rho_{\alpha} (\omega) e^{-i \omega (t-t^{\prime})},
\nonumber\\ 
g^{<}_{\alpha}(t-t^{\prime}) &=& i \int \frac{d \omega}{2\pi} 
f_{\alpha}(\omega) \rho_{\alpha}(\omega) e^{-i \omega (t-t^{\prime})}.
\label{e22}
\end{eqnarray}
The Green functions $G^R_{1,1}(t, t^{\prime})$ and $G^<_{1,1}(t, t^{\prime})$
correspond to the ring coupled to the system $\alpha$. They satisfy the
following equations
\begin{eqnarray}
G^R_{1,1}(t,t^{\prime})&=&
g^R_{1,1}(t,t^{\prime}) +\int dt_1 dt_2 G^R_{1,1}(t,t_1) \Sigma_1^R(t_1,-t_2)
g^R_{1,1}(t_2,t^{\prime}) ,\nonumber\\
G^<_{1,1}(t,t^{\prime})&=& \int dt_1 dt_2 G^R_{1,1}(t,t_1) \Sigma^<_1(t_1-t_2)
G^A_{1,1}(t_2,t^{\prime}),
\label{e23}
\end{eqnarray}
where $g^R_{1,1}(t,t^{\prime})$ is the retarded Green function for the
uncoupled ring penetrated by the magnetic field. The self-energies 
\begin{eqnarray}
\Sigma_1^R(t-t^{\prime})&=& |T_{1\alpha}|^2 g^R_{\alpha} (t-t^{\prime}),
\nonumber \\
\Sigma_1^<(t-t^{\prime})&=& |T_{1\alpha}|^2 g^<_{\alpha} (t-t^{\prime})
\label{e24}
\end{eqnarray}
account for the corrections due to the escape to the leads \cite{past,jau}.
This set of equations provide the  exact solution for the time-dependent
current $J_{1\alpha}(t)$. 

We now introduce one simple model to describe the system  $\alpha$.
To simplify the numerical procedure, we assume that the ensuing
electronic band has a large bandwidth and a constant density of states
 $\rho_{\alpha} (\omega) = \Gamma$.  Thus, 
$g^R_{\alpha}(t-t^{\prime}) \sim -i \Gamma \delta(t-t^{\prime})$ and
 the  following expression for the time-dependent current is obtained
\begin{equation}
J_{1\alpha}(t)=\frac{2e}{\hbar} \sigma \int \frac{d \omega}{\pi}
f_{\alpha}(\omega) (Im[G^R_{1,1}(t,\omega)]+ \sigma |G^R_{1,1}(t,\omega)|^2),
\label{e25}
\end{equation}
where $\sigma= |T_{1\alpha}|^2 \Gamma$, with 
$\Sigma_1^R(t-t')=i \sigma \delta(t-t')$, 
while the Fourier transform for 
 the retarded Green function is defined as,
\begin{equation}
G^R_{m,n}(t,\omega)=\int_{-\infty}^t dt^{\prime} 
e^{i (\omega+i \eta)(t-t^{\prime})}G^R_{m,n}(t,t^{\prime}),
\label{fourier}
\end{equation}
with $\eta=0^+$.

We remark now some salient issues related with expression (\ref{e25}).
The function $\rho_l(t,\omega)=-2 Im[G^R_{l,l}(t,\omega)]$ defines
a generalized time-dependent spectral function for site $l$, while the function
$N_l(t,\omega)=2 \sigma |G^R_{l,1}(t,\omega)|^2 $ defines the occupation of 
this site at time $t$, since
\begin{equation}
N_l(t)= -i G^<_{ll}(t,t)= 
\int \frac{d\omega}{2 \pi} f_{\alpha}(\omega) N_l(t,\omega),
\label{ocupa}
\end{equation}
where $N_l(t)=\langle c^{\dagger}_l c_l \rangle$ is the mean number of 
particles at site $l$.
In an equilibrium system or in a steady regime where  
$G^R_{l,l}(t,\omega)$ does not have an explicit time dependence,
 it is easy to prove that 
$N_l(\omega)=\rho_l(\omega)$. However, in a non-equilibrium time-dependent 
regime the local Green
function $G^R_{l,l}(t,\omega)$ depends explicitly on time and
 can develop  a complex structure
as a function of $\omega$. Therefore, 
the two functions are in general different.
This property is crucial for the physical effects we aim to address. In 
particular,
the fact that in a stationary situation these two functions coincide, cause 
the current $J_{1\alpha}(t)$
 to vanish. Instead, in a non-equilibrium regime, an ac
current can flow through the lead. As we shall see, 
this current exhibits a periodic behavior which
  is originated by the existence of Bloch oscillations
in the ring.

To complete the theoretical description of the system, we define the
current flowing through the bond between sites $l$ and $l+1$
along the ring
\begin{equation}
J_{l,l+1}= \frac{2e}{\hbar} Re[ T e^{-i \phi t} G^{<}_{l,l+1}(t,t)],
\label{curl}
\end{equation}
where 
\begin{equation}
G^{<}_{l,l+1}(t,t) =  i \int \frac{d \omega}{\pi} 
\sigma G^R_{l,1}(t,\omega) [G^R_{l+1,1}(t,\omega)]^*,
\label{curl1}
\end{equation}
with
\begin{equation}
G^{R}_{m,n}(t,t^{\prime}) = g^{R}_{m,n}(t,t^{\prime})
+ i \int d t_1 G^{R}_{m,1}(t,t_1) \sigma  g^{R}_{1,n}(t_1,t^{\prime}).
\label{dyson}
\end{equation}

From the practical point of view, the problem reduces to the calculation
of the retarded Green functions $g^R_{m,n}(t,t^{\prime})$ and 
$G^R_{m,n}(t,t^{\prime})$ corresponding the loop penetrated 
by the magnetic field unconected and coupled to the external reservoir, 
respectively.
This is the subject of Sec. III.

\section{Calculation of the retarded Green functions}

\subsection{The metallic ring threaded by a flux $\Phi_M(t)=\Phi t$}

We are interested in calculating the retarded Green function 
$g^R_{m,n}(t,t^{\prime})$, where $m,n$ label the $m$-th and 
$n$-th sites of the ring, respectively. The usual procedure to compute
this quantity is to take advantage of the 
translational invariance of the system \cite{four}. In this subsection, we
shall discuss an alternative way to obtain {\em exactly} 
$g^R_{m,n}(t,t^{\prime})$.  
As we shall see, this route leads to a more
transparent physical picture for the relevant energy scales of the problem 
while it has the additional advantage of being adequate 
 to solve more general problems without translational
invariance. 
We shall see in the next subsection that 
this procedure also provides a very convenient framework
to treat the coupling of the ring with the external lead and 
reservoir.

The Dyson equation for the retarded Green function leads to
the following set of coupled differential equations
  \begin{equation}
-i \hbar \frac{\partial}{\partial t^{\prime}}  
g^R_{m,n}(t,t^{\prime}) +  
T e^{i \phi t^{\prime}} g^R_{m,n+1}(t,t^{\prime}) 
+T e^{-i\phi t^{\prime}} g^R_{m,n-1}(t,t^{\prime} ) 
=\delta(t-t^{\prime})\delta_{m,n}. 
\label{e3}
\end{equation}
Our strategy is to perform a gauge transformation in order to
translate the original problem into an equivalent one where the explicit
time dependence appears only in the matrix elements involving the link 
between the first and the last sites of the
ring. 
By defining the  gauge transformation
\begin{equation}
 c_n=\exp[i n \phi t] \overline{c}_n,
\label{gauge}
\end{equation}
the Green function transforms as
\begin{equation} 
g^R_{m,n}(t, t^{\prime})=\exp[i \phi (m t - n t^{\prime})]  
\overline{g}^R_{m,n}(t, t^{\prime}). 
\label{e4}
\end{equation} 
From (\ref{e3}), it is found that $\overline{g}^R_{m,n}(t, t^{\prime})$ 
satisfies
\begin{eqnarray} 
& &-i \hbar \frac{\partial}{\partial t^{\prime}} 
\overline{g}^R_{m,n}(t, t^{\prime}) -
\hbar \phi n   \overline{g}^R_{m,n}(t,
t^{\prime}) +
 T \overline{g}^R_{m,n+1}(t, t^{\prime})+ 
T \overline{g}^R_{m,n-1}(t, t^{\prime})\nonumber\\
& &=\delta(t,t^{\prime})\delta_{m,n},
\;\;\;\;\;\;n\neq 1,N  \nonumber \\ 
& &-i \hbar \frac{\partial}{\partial t^{\prime}} \overline{g}^R_{m,1}(t,
t^{\prime})-  \hbar \phi \overline{g}^R_{m,1}(t, t^{\prime}) +  
T \overline{g}^R_{m,2}(t, t^{\prime})+ 
T_{N1}(t^{\prime})
\overline{g}^R_{m,N}(t,t^{\prime})\nonumber \\
& &=\delta(t,t^{\prime})\delta_{m,1},
\;\;\;\;\;\;n= 1  \nonumber \\ 
& &-i \hbar \frac{\partial}{\partial t^{\prime}} \overline{g}^R_{m,N}
(t, t^{\prime}) 
-\hbar \frac{\Phi}{\Phi_0} \overline{g}^R_{m,N}(t, t^{\prime}) 
+  
T _{1N}(t^{\prime})    \overline{g}^R_{m,1}(t, t^{\prime})+ 
T \overline{g}^R_{m,N-1}(t, t^{\prime})\nonumber \\
& &=\delta(t,t^{\prime})\delta_{m,N}.
\;\;\;\;\;\;n= N, 
\label{e5}
\end{eqnarray} 
where the only matrix elements that exhibit an explicit time dependence are
\begin{equation}
T_{1N}(t)=\exp[i\frac{\Phi}{\Phi_0} t] T,
\label{t1n}
\end{equation}
 with $[T_{N1}(t)]^*=T_{1N}(t)$.

The above set can be interpreted as the Dyson equation for the retarded
 Green function corresponding to the problem of a chain with $N$ sites 
of tight-binding
electrons in the presence of a  dc voltage
\begin{equation}
V_l= \hbar l \phi.
\label{bias}
\end{equation}
The chain is  closed at its ends by a time dependent link $T_{1N}(t)$.  
We formally treat this system as indicated in the scheme of Fig. 2, by 
separating the ensuing hamiltonian in an ``unperturbed'' part
that describes an open tight-binding chain,
\begin{equation} 
\overline{H}^0\;=\; 
-T \sum_{l=1}^{N-1}  
(\overline{c}^{\dagger}_l  
\overline{c}_{l+1} + h.c) 
+ \sum_{l=1}^{N} V_l \overline{c}^{\dagger}_l  
\overline{c}_l , 
\label{e6} 
\end{equation} 
and a time-dependent ``perturbation'' 
\begin{equation}
\overline{H}_{1N}(t)=-T_{1N}(t) \overline{c}^{\dagger}_{1} 
\overline{c}_{N}
- T_{N1}(t) \overline{c}^{\dagger}_{N} \overline{c}_{1},
\label{e7}
\end{equation}
which describes  a time-dependent hopping between the first and the last
sites of the chain.
Within this picture, the integral form for the Dyson equation 
reads 
\begin{eqnarray} 
\overline{g}^R_{m,n}(t,t')&=& 
g^0_{m,n}(t,t') -\int_{t'}^t dt_1 
\overline{g}^R_{m,N}(t,t_1) T_{N1}(t_1) g^0_{1,n}(t_1,t')  
\nonumber \\ 
&-& \int_{t'}^t dt_1 
\overline{g}^R_{m,1}(t,t_1) T_{1N}(t_1) g^0_{N,n}(t_1,t'), 
\label{e8} 
\end{eqnarray} 
where the  matrix elements $T_{N1}(t)$ and  $T_{1N}(t)$ 
define time-dependent instantaneous self-energies. 
The ``unperturbed'' Green function  $g^0_{m,n}(t,t') \equiv
g^0_{m,n}(t-t')$ is an  {\em equilibrium} Green function 
corresponding to the hamiltonian $\overline{H}_0$. 
Although there is no trivial analytical expression for $g^0_{m,n}(t-t')$, it
can be easily obtained in a finite-size system from the numerical solution of
hamiltonian (\ref{e6}). 
In terms of the eigenvalues $E_{\nu}$ and eigenvectors
$|\nu \rangle= \sum_l A^{\nu}_l |l\rangle$ of $\overline{H}_0$,  
it can be written as
\begin{equation}
g^0_{m,n}(t-t')=-i \Theta (t-t^{\prime})
\sum_{\nu=1}^N A^{\nu}_{m} A^{\nu}_{n}
\exp[- \frac{i}{\hbar} E_{\nu} (t-t^{\prime})].
\label{e9}
\end{equation}

After performing in Eqs. (\ref{e8}) the Fourier transform defined in 
(\ref{fourier}),
we obtain for $n=1$ 
\begin{eqnarray} 
 \overline{g}^R_{m,1}(t,\omega)+\overline{g}^R_{m,N}(t,\omega+\frac{\Phi}{\Phi_0}) 
T_{N1}(t) g^0_{1,1}(\omega)
+ \overline{g}^R_{m,1}(t,\omega-\frac{\Phi}{\Phi_0}) 
T_{1N}(t) g^0_{N,1}(\omega)&=&g^0_{m,1}(\omega)\nonumber \\ 
\overline{g}^R_{m,N}(t,\omega)+\overline{g}^R_{m,N}(t,\omega+\frac{\Phi}{\Phi_0}) 
T_{N1}(t) g^0_{1,N}(\omega)  
+ \overline{g}^R_{m,1}(t,\omega-\frac{\Phi}{\Phi_0}) 
T_{1N}(t) g^0_{N,N}(\omega)&=&g^0_{m,N}(\omega). 
\label{e11} 
\end{eqnarray}   
For each time $t$, the solution of the above set of linear equations
provides the complete exact solution of the problem. 
The Green function 
for arbitrary $m,n$ is obtained from
\begin{equation}
\overline{g}^R_{m,n}(t,\omega)=
g^0_{m,n}(t,\omega)-
\overline{g}^R_{m,N}(t,\omega+\frac{\Phi}{\Phi_0})T_{N1}(t)g^0_{1,n}(\omega)-
\overline{g}^R_{m,1}(t,\omega-\frac{\Phi}{\Phi_0})T_{1N}(t)g^0_{N,n}(\omega),
\label{e12}
\end{equation}
while the Fourier transform of the retarded Green function
of the original problem is
\begin{equation}
g_{m,n}(t,\omega)=\exp[i \phi (m-n) t]
\overline{g}^R_{m,n}(t,\omega-n \phi).
\label{e13}
\end{equation}

\subsection{The ring coupled to the particle reservoir}

We now develop a computational strategy to obtain the retarded
Green function $G^R_{m,n}(t,t^{\prime})$ for the coupled system.
The differential form of the Dyson equation (\ref{dyson}) for this
Green function reads
\begin{eqnarray}
& &-i\hbar \frac{\partial }{\partial t^{\prime}} G^R_{m,n}(t,t^{\prime})
-i \sigma \delta_{n,1} G^R_{m,n} + T e^{-i \phi t^{\prime}}
G^R_{m,n+1}(t,t^{\prime}) \nonumber\\
& &+ T e^{i \phi t^{\prime}}
G^R_{m,n-1}(t,t^{\prime}) =\delta_{m,n} \delta(t-t^{\prime}).
\label{ec1}
\end{eqnarray}

As in the previous subsection, we perform the gauge transformation
(\ref{gauge}) in order to obtain a set of equations with a
reduced number of time-dependent matrix elements. The Green
function transforms as 
\begin{equation}
G^{R}_{m,n} (t, t^{\prime})=\exp[i \phi (m t - n t^{\prime})]  
\overline{G}^R_{m,n}(t, t^{\prime}) 
\label{ec2}
\end{equation} 
and $\overline{G}^R_{m,n}(t, t^{\prime})$ satisfies the following equations
\begin{eqnarray}
& &-i \hbar \frac{\partial }{\partial t^{\prime}} 
\overline{G}^R_{m,n}(t,t^{\prime})
- \hbar \phi n \overline{G}^R_{m,n}(t,t^{\prime})
+T \overline{G}^R_{m,n+1}(t,t^{\prime})+
T \overline{G}^R_{m,n-1}(t,t^{\prime})\nonumber\\
& &=\delta_{m,n}\delta(t-t^{\prime})
\;\;\;\;\;n \neq 1, N\nonumber\\
& &-i \hbar \frac{\partial }{\partial t^{\prime}} 
\overline{G}^R_{m,1}(t,t^{\prime})
- \hbar \phi  \overline{G}^R_{m,1}(t,t^{\prime})
- i \sigma \overline{G}^R_{m,1}(t,t^{\prime})
+T \overline{G}^R_{m,2}(t,t^{\prime})+
T_{N,1}(t^{\prime})
 \overline{G}^R_{m,N}(t,t^{\prime})\nonumber \\
& &=\delta_{m,1}\delta(t-t^{\prime})\nonumber\\
& &-i \hbar \frac{\partial }{\partial t^{\prime}} 
\overline{G}^R_{m,N}(t,t^{\prime})
- \hbar \frac{\Phi}{\Phi_0}  \overline{G}^R_{m,N}(t,t^{\prime})
+T_{1,N}(t^{\prime}) \overline{G}^R_{m,1}(t,t^{\prime})+
T \overline{G}^R_{m,N-1}(t,t^{\prime})\nonumber \\
& &=\delta_{m,N}
\delta(t-t^{\prime})
\label{ec3}
\end{eqnarray}

The above set is formally identical to 
the Dyson equation of the problem illustrated in Fig. 3. 
This consists in an open tight-binding chain of $N-1$ sites labeled 
by $l=2,\ldots,N$ with potentials $V_l$ (see (\ref{bias})), plus an 
additional site
labeled by $l=1$ (indicated with a square in Fig. 3). The latter represents
a site of the ring coupled to the lead and reservoir. It has a
local potential $V_1$ and  
a self-energy correction $i \sigma$ due to the coupling to the 
system $\alpha$. 
This site is connected to one end of the chain through a hopping $T$
and to the other end, through the time-dependent hopping 
$T_{1N}(t)$ (\ref{t1n}).  
The integral form of the Dyson equation results,
\begin{eqnarray}
\overline{G}^R_{m,n}(t,t^{\prime})&=&
g^0_{m,n}(t,t^{\prime})-
\int d\omega dt_1 \overline{G}^R_{m,N}(t,t_1)T_{N1}(t_1)
g^0_{1,n}(t_1,t^{\prime})-\nonumber\\
& &\int d\omega dt_1 \overline{G}^R_{m,1}(t,t_1)T_{1N}(t_1)
g^0_{N,n}(t_1,t^{\prime})-
T\int dt_1 \overline{G}^R_{m,1}(t,t_1) g^0_{2,n}(t_1,t^{\prime})
- \nonumber \\
& &T \int dt_1 \overline{G}^R_{m,2}(t,t_1) g^0_{1,n}(t_1,t^{\prime}),
\label{ec4}
\end{eqnarray}
which by recourse to the  Fourier transform (\ref{fourier}) leads to 
the following set
\begin{eqnarray}
\overline{G}^R_{m,1}(t,\omega)&=& g^0_{m,1}(\omega)- 
T e^{-i \frac{\Phi}{\Phi_0} t} 
\overline{G}^R_{m,N}(t,\omega+\hbar \frac{\Phi}{\Phi_0}) g^0_{1,1}(\omega)
-T \overline{G}^R_{m,2}(t,\omega) g^0_{1,1}(\omega)
\nonumber\\ 
\overline{G}^R_{m,N}(t,\omega)&=& g^0_{m,N}(\omega)- 
T e^{i \frac{\Phi}{\Phi_0} t} 
\overline{G}^R_{m,1}(t,\omega-\hbar \frac{\Phi}{\Phi_0}) g^0_{N,N}(\omega)
-T \overline{G}^R_{m,1}(t,\omega) g^0_{2,N}(\omega)
\nonumber\\ 
\overline{G}^R_{m,2}(t,\omega)&=& g^0_{m,2}(\omega)- 
T e^{i \frac{\Phi}{\Phi_0} t} 
\overline{G}^R_{m,1}(t,\omega-\hbar \frac{\Phi}{\Phi_0}) g^0_{N,2}(\omega)
-T \overline{G}^R_{m,1}(t,\omega) g^0_{2,2}(\omega),
\label{ec5}
\end{eqnarray}
where $g^0_{m,n}(\omega)$ with $m,n=2,\ldots,N$ 
is the Green function of the chain  
formed with sites $2,\ldots,N$ with a bias $V_l$,
while
$g^0_{1,1}= 1/(\omega -\phi +i \sigma)$ is the 
Green function for the remaining site 
in interaction with system $\alpha$.

After some rather simple algebra, it is easy to see that
Eqs. (\ref{ec5}) can be reduced to the set (\ref{e11})   
by replacing in the latter $\overline{g}^R_{m,n}(t,\omega)$
by $\overline{G}^R_{m,n}(t,\omega)$
and $g^0_{m,n}(t,\omega)$
by $G^0_{m,n}(t,\omega)$, being
\begin{equation}
G^0_{m,n}(\omega)=g^0_{m,n}(\omega)-  T G^0_{m,1}(\omega) g^0_{2,n} -
T G^0_{m,2}(\omega) g^0_{1,n}(\omega).
\label{ec6}
\end{equation}
A scheme of the underlying physical processes related with this equation
is given in Fig. 3b. It corresponds to the coupling of the open chain with the
site $l=1$ by means of a single hopping element $T$.
Equations (\ref{e11}) correspond to the closing of the resulting $N$-site
chain by means of a time-dependent hopping process (\ref{e7}) between its ends.

In conclusion, the {\em exact } Green function for the loop penetrated by
the time-dependent magnetic field, as an isolated system or as a system 
coupled to
an external particle reservoir, is given by the solution of Eqs.
(\ref{e11}), provided that a suitable expression
for the ``unpperturbed'' Green function, $g^0_{m,n}(t,\omega)$ or 
$G^0_{m,n}(t,\omega)$ correspondingly, is given.
We must emphasize that although in both cases the final set of equations is 
formally identical, physical processes of different nature are enclosed in 
the ``unperturbed'' Green
function. In particular in the case considered in this subsection, the effect 
of the lead represented by the self-energy $i \sigma$ introduces 
inelastic 
scattering events that
produce loss of coherence in the propagation of the wave function through 
the ring and dissipation.   

The imaginary part of the
local Green function defines a generalized time-dependent density of states
$\rho_l(t,\omega)= -2 Im[ G^R_{l,l}(t,\omega)]$,
in terms of which the local Green function can be written as
\begin{equation}
G^R_{l,l}(t,\omega)=\int \frac{d\omega}{2 \pi} \frac{\rho_{l}(t,\omega)}
{\omega-\omega^{\prime}+i \eta}.
\end{equation}
This quantity is in general expected to be a periodic function on time, 
as suggested by the structure of the set (\ref{e11}). However, we shall 
later see that in some situations it can also become independent of time.

\section{Results}

\subsection{Dynamical regimes}
We first consider the case introduced in Section IIIA, which
corresponds to the ring under the influence of the time-varying magnetic flux,
but disconnected from the external wire and particle reservoir.
Our aim is to understand first the dynamical behavior of this system, 
free from the effects of dissipation introduced by the coupling to the 
external reservoir.
   
An important issue is that if $g^0_{1,N}(\omega)=g^0_{N,1}(\omega)=0$,
the equations in (\ref{e11}) can be decoupled. In that case, the solution 
becomes stationary, i.e. does not have any dependence on time and reads
\begin{equation}
\overline{g}^R_{m,n}(t,\omega)
\equiv \overline{g}^R_{m,n}(\omega)=
\frac{g^0_{m,n}(\omega)}{1-T^2 g^0_{11}(\omega)g^0_{NN}(\omega+\Phi/\Phi_0)}.
\label{ed20}
\end{equation} 
The latter property implies that the local Green functions of the original
problem $g_{l,l}(t,\omega)$ become independent of time, while the non-local 
ones exhibit a trivial time dependence introduced by the gauge transformation 
(\ref{e13}).

In order to find out the physical conditions in which
this stationary solution is expected, we study the behavior of 
$g^0_{1,N}(\omega)$.
 The relevant parameters are
the magnitude $\hbar \Phi/\Phi_0$ relative to the bandwidth of the
tight-binding chain $W=4T$ and the size of the system $N$.
 We numerically diagonalized the hamiltonian (\ref{e6}) for different ratios
$\hbar \Phi/\Phi_0 W$ and studied systems with sizes up to $N=800$ sites. 
Typical curves for the imaginary part  
\begin{equation}
Im[g^0_{1,N}(\omega)]= -\pi 
\sum_{\nu=1}^{N} A^{\nu}_1 A^{\nu}_N \delta(\omega - E_\nu ).
\label{ed21}
\end{equation}
for a chain with $N=20$ sites are shown in Fig. 4.   
It is clear that for all $\omega$, the magnitude of this quantity
is significantly larger in the cases where $\hbar \Phi/\Phi_0 < W$.
 
To be able to perform a more quantitative analysis of the  behavior 
of $g^0_{1,N}(\omega)$ as a
function of $N$, we consider the function $g_{max}(N)$, which is defined
as the maximum value achieved by $|Im[g^0_{1L}(\omega)]|$  for a given size
$N$. Plots of $g_{max}(N)$ as a function of $N$ reveal that there is
a change in the scaling of this quantity as  $\hbar \Phi / \Phi_0 W$ varies.
The case $\hbar \Phi \leq \Phi_0 W$ is illustrated in Fig. 5. We found that the
scaling law is 
\begin{equation}
g_{max}(N) \propto \frac{1}{N^{\gamma}}, \;\;\;\; \hbar \Phi/\Phi_0 \leq W.
\label{ed22}
\end{equation}
The data of Fig. 5 can be adequately fitted with $\gamma \sim 1$ for 
$\hbar \Phi/\Phi_0 W =0,0.25,0.5,0.75$ and  $\gamma \sim 1.4$ for
 $\hbar \Phi/\Phi_0 W =1$. 
The opposite situation, where $\hbar \Phi/\Phi_0 > W$ is shown in Fig. 6.
We found that in this case, the scaling law is
\begin{equation}
g_{max}(N) \propto\ e^{-\frac{N}{L_{\alpha}}}, \;\;\;\; \hbar \Phi/\Phi_0 > W.
\label{ed23}
\end{equation}
 By fitting the numerical data it is obtained
$L_{\alpha} \sim 8$ for $\hbar \Phi /\Phi_0 W = 1.25$ and 
$L_{\alpha} \sim 4$ for $\hbar \Phi / \Phi_0 W = 1.5$.

The existence of two kinds of behaviors in the scaling of $g^0_{1,N}(\omega)$
 has a rather simple explanation. We
recall that  this Green function corresponds to a system of
 electrons in  an open tight-binding chain with a bias $\hbar
\Phi/\Phi_0 $ between its edges. The different scaling laws
 result from a concurrence between
the kinetic  energy of the electrons in the chain and the strength of the
bias.  In fact, the Green function  $g^0_{1,N}(t-t')$ is by definition
related with overlaps of the form
$ \langle \Psi_1(t)|\Psi^{\prime}_L(t^{\prime}) \rangle $, where
$|\Psi_{l}(t) \rangle$ and $|\Psi^{\prime}_{l}(t) \rangle$
are the wave functions of the system when an electron is created and destroyed,
respectively, in site $l$ at time $t$. This quantity is thus a measure for the
ability  of an electron to move from the first to the last site of the chain
in a time $t-t^{\prime}$. In the free chain, the probability of such a movement
decreases as $1/N$ as the size of the system increases, independently of how
large the time it takes is assumed to be.  
Our numerical analysis indicates that this is also the case in the presence
of a bias, provided that the strength of the bias is not larger than the
bandwidth associated with the kinetic energy of the electrons. 
As the bias increases and overcomes the value of the bandwidth, this
probability becomes smaller and decreases exponentially with the size of the
system.    

Therefore, for magnetic fields varying at a rate such that
$\hbar \Phi/\Phi_0 \leq W$, 
the Green functions $\overline{g}^R_{m,n}(t,\omega)$ 
tend to the steady solution (\ref{ed20}) as $N^{-\gamma}$ when
the size of the system increases. 
Hence, for finite size systems, time dependent features are  
clearly observable. 
As a function of time, the functions $\overline{g}^R_{m,n}(t,\omega)$ 
oscillate with the period $\tau=2\pi\Phi_0/\Phi$. The latter being the
period of the Bloch oscillations 
due to the induced electric field of magnitude $F=\Phi/cL$.
Instead, for rapid variations of the magnetic field such that
$\hbar \Phi/\Phi_0 > W$, the Green 
functions tend to (\ref{ed20}) exponentially fast as the length of the 
circumference 
increases. As a consequence, within this regime,
time dependent features can be observed only
in very small rings of lengths smaller than $L_{\alpha}$.

We close this subsection with some comments about the behavior introduced by 
inelastic scattering processes in the dynamical behavior above mentioned. 
We have 
shown in section IIIB
that the Green function for the coupled ring is also given by the solution of 
Eqs. (\ref{e11})
with the replacement of $g^0_{m,n}(\omega)$ by the function $G^0_{m,n}(\omega)$
which contains the corrections due to the escape to the leads. Thus, the 
equations in (\ref{e11}) are decoupled and a stationary solution 
\begin{equation}
\overline{G}^R_{m,n}(\omega)
\equiv \overline{G}^R_{m,n}(\omega)=
\frac{G^0_{m,n}(\omega)}{1-T^2 G^0_{11}(\omega)G^0_{NN}(\omega+\Phi/\Phi_0)}.
\label{ed25}
\end{equation} 
exists also in this case when the function
$G^0_{1L}(\omega)$ becomes vanishingly small. As before, this quantity
is related with the probability of an electron to travel from the
first to the last sites of the chain. In this case, the new ingredient
is the existence of a typical inelastic scattering length   
$L_{in}$ beyond which the coherence in the wave packet propagation is loss.
The latter is related to the imaginary part of the self-energy 
through $\sigma \propto 1/L_{in}$ which generates damping and
produces an exponential 
decay $\sim \exp[-\alpha L/L_{in}]$
in  the Green functions $G^0_{m,n}(\omega)$  \cite{damato}. 
Therefore, for rings with circumferences $L<L_{in}$  
the same two dynamical regimes defined for the disconnected ring can be
distinguished. Instead, as soon as $L>L_{in}$,
inelastic scattering due to the presence of the external reservoir
 dominate the physical behavior
of the system, the Green functions coincide with 
the steady solution (\ref{ed20}) and no time dependent 
features are observed. 
An example of significant time-dependent behavior
is shown in Fig. 7 where  
typical shapes of the local generalized density of states 
$\rho_1(t,\omega)$
for a ring of $N=20$ sites with $\sigma=0.025W$  
at different times are depicted. 
As a function of $\omega$, $\rho_1(t,\omega)$
 can be either positive or negative as a consequence
of the far from equilibrium nature of this regime. The sum rule
\begin{equation}
\int \frac{d\omega}{2 \pi}  \rho_1(t,\omega) =1
\label{ed26}
\end{equation}
is, however, satisfied, justifying the name of density of states.

\subsection{The current through the lead}
Although a classical textbook analysis indicates that the current
through the single lead coupling the ring to the particle reservoir
 should vanish, within the time-dependent
regime defined in the previous subsection, there is an
ac contribution  which could be measured in a real experiment. 

The time-dependent regime is characterized by an imaginary 
part of the self energy $\sigma$ leading to an inelastic scattering
length $L_{in}$ larger than the size of the ring and a variation
of the magnetic field satisfying the condition $\hbar \Phi/\Phi_0 leq
W$.
The propagation of the
wave packet maintains the coherence, the Green functions  
$g_{1,1}(t, \omega)$ and  $G_{1,1}(t, \omega)$ oscillate 
with the period $\tau=2\pi\Phi_0/\Phi$ of Bloch oscillations driven by the 
induced electric field, and so does the current through the lead.
Examples for a ring with $N=20$ sites 
are shown in Fig. 8 where $J_{1 \alpha}(t)$, obtained from 
(\ref{e25}), is shown for two
different values of $\Phi$. The imaginary part of the self-energy 
is taken to be $\sigma=0.2W$.
The physical picture that emerges is that 
the induced electric field produces dramatic changes
in the generalized time-dependent density of states in the ring
$\rho_1(t,\omega)$.
This quantity is periodic in time and eventhough it satisfies the
sum rule (\ref{ed26}), it can achieve positive as well as negative
values as a function of $\omega$ (see Fig. 7). In particular,
this function does not satisfy the 
property of being equal to the occupation function 
$N_1(t,\omega)$. As a consequence the integral (\ref{e25})
does not vanish in general but displays oscillations
with zero mean value as a function of time. 
A somehow similar phenomenon in the context of optical
properties of semiconductors is the so called Franz-Keldysh effect \cite{fk}.
In that case, the changes in the density of states due to the
influence of time-dependent fields cause the absorption coefficient to become
finite for photon energies bellow the band-edge.
 
Interestingly, due to the symmetry of the spectrum (see Fig. 7)
the current $J_{1 \alpha} (t)$ is an antisymmetric function of the 
chemical potential $\mu$. This property is due to the particular
geometry of the system we are considering. Any spectral asymmetry,
introduced, for example, by the presence of barriers and/or disorder
in the ring should remove this feature. 
            
The effect of inelastic scattering is illustrated in Fig. 9, where    
$J_{1\alpha}$ is shown for several values of $\sigma$.
Inelastic scattering introduces damping and $L_{in}$
decreases as $\sigma$ increases. 
In terms of the spectrum, the peaks of
$\rho_1(t,\omega)$ get wider and, for very large $\sigma$, the
time dependent features tend to be wiped out. On the other hand,
very weak inelastic scattering, which occurs for example for
 a very small value of $T_{1\alpha}$, tends to forbid the quantum 
tunneling between the ring and the reservoir. Hence, the amplitude
for the oscillations of $J_{1\alpha}$ is vanishingly small for
small $\sigma$, tends to decrease for very large $\sigma$ and
attains a  maximum  in between.  

\subsection{The current along the ring}
The resistive effect caused by inelastic scattering in metallic loops
was studied some time ago by Landauer and B\"uttiker \cite{lb},
who proposed a phenomenological equation of motion
to analyze the relaxation processes in a ring penetrated
by a time-dependent magnetic flux. A dc component in the electric current
along the ring was found to increase as inelastic scattering is 
introduced and to decrease as inelastic scattering
 becomes sufficiently intense. This analysis was later supplemented by
B\"uttiker \cite{b}, who proposed a single lead connecting the
ring with an external reservoir like in Fig. 1, as a simple device
to generate a resistance in the metallic loop. 

In this section we show that the theoretical approach adopted in the
present work is also appropriate to investigate the effect of inelastic
scattering in the current along the ring. 
Results for the current
(\ref{curl}) through the bond between the sites $l=1,2$
of the ring are shown in Fig. 10 for a system with $N=20$ sites.
We consider parameters within the time-dependent regime.
The imaginary part of the self-energy is assumed to be $\sigma=0.2W$.
As a function of time, the current $J_{12}(t)$ displays oscillations
with the period $\tau=2\pi\Phi_0/\Phi$ of Bloch oscillations.
Note that the factor $e^{-i\phi t}$ in (\ref{curl}) exactly
cancels the time-dependent exponential introduced by the
gauge transformation (\ref{e13}). Thus, the oscillations
of the current $J_{l,l+1}$ result as a 
consequence of the time-dependent behavior of the Green functions
$\overline{G}^R_{m,n}(t,\omega)$ obtained from the solution of Eqs.
 (\ref{e11}) and
disappear in the steady regime. The remarkable feature is the 
existence of a dc component in the current, which depends on the
chemical potential of the reservoir, as well as on the rate
$\hbar \Phi/\Phi_0$ for fixed $\sigma$.
The effect of $\sigma$ on the magnitude of the dc component is
shown in Fig. 11, where it is observed
the behavior predicted in Ref \cite{lb}, namely 
the magnitude of this component is vanishes
for small as well as for very large
$\sigma$ and has a maximum for intermediate strength
of dissipation. 

To close this section, we show the behavior of the mean 
occupation number $N_l(t)$ along the ring. The latter is calculated
from (\ref{ocupa}) for a ring with $N=20$ sites. Results
are shown in Fig. 12 for some values of $l$. As a function of time
it displays oscillations with period $\tau$ around the mean value
\begin{equation}
\langle N_l \rangle=\frac{1}{\tau}\int dt N_l(t),
\end{equation}
that depends strongly on the position $l$, as shown in Fig. 13.
The behavior of $\langle N_l \rangle$ is a measure of the effective
mean potential along the chain. The latter exhibits a significant
drop near the site $l=1$ where the lead is connected, which becomes
more abrupt as the strength of the inelastic scattering introduced
by $\sigma$ increases.  

\section{Summary and conclusions}
We have studied the transport properties of a ring
threaded by a  magnetic
flux with a linear dependence on time $\Phi_M(t)=\Phi t$.
We considered an experimental setup to investigate the
transport properties of this system, which consists
in a single lead connecting the ring to an external
particle reservoir. We employed
a theoretical approach based in Baym-Kadanoff-Keldysh
non-equilibrium Green functions. We derived a system
of linear equations to calculate in real space the retarded Green
function of the disconnected ring as well as for the ring 
coupled to the particle reservoir. 

This method provides a transparent picture for the different energy scales
governing the dynamics of the system. In particular,
we identified two kind of behaviors  in the Green functions
of the uncoupled ring, depending on
the ratio $\hbar \Phi/\Phi_0 W$. For magnetic fields
varying at a rate such that $\hbar \Phi/\Phi_0 \leq W$,
the time dependent features are important in the
behavior of the Green functions of finite rings
and tend to disappear as $N^{-\gamma}$ when the size of the system increases.
Instead, for $\hbar \Phi/\Phi_0 > W$, the Green functions tend
exponentially fast to a static solution as the size of the system
increases. These two kinds of behaviors manifest themselves in the transport
properties of the system when the inelastic scattering length is larger
than the perimeter of the ring. Within the time-dependent regime,
the current through the lead as well as the current along the
ring and the charge inside it, display oscillations with the period 
$\tau=2\pi\Phi_0/\Phi$ of Bloch oscillations
driven by the induced electric field. In the situation of 
a rapid magnetic field variation satisfying $\hbar \Phi/\Phi_0 > W$, 
no current is measured in the lead and no oscillatory behavior is 
observed  except
for extremely small rings. We stress that this effect has its grounds
in the concurrence between the kinetic energy of the electrons 
and the strength of the dc voltage due to the induced electric field,
being independent of the loss of coherence introduced by inelastic scattering.

We finally analyzed the effect of inelastic scattering caused by the coupling
with the external reservoir in the time-dependent transport. 
We found that the amplitude of the oscillations 
of the current through the lead as well as the dc component of the    
current along the ring, tend to vanish in the two limiting situations
of very weak and very strong inelastic scattering, attaining a maximum
for intermediate strengths.
 
On the technical side, we find that the present approach provides
an useful theoretical tool 
to investigate the effects of time
dependent magnetic fluxes and induced electric fields
in other interesting systems of larger complexity,
like metallic loops with disorder, rings with 
inserted barriers and quantum dots and one dimensional systems with
electronic correlations.

\section{Acknowledgements.}
L. A. is supported by CONICET and Fundaci\'on Antorchas.
Support from the AvH foundation at the
first stage of this work is acknowledged.
The author thanks G. Chiappe and H. Pastawski
for stimulating discussions.

\begin{figure}
\epsfxsize=3.5in
\epsffile{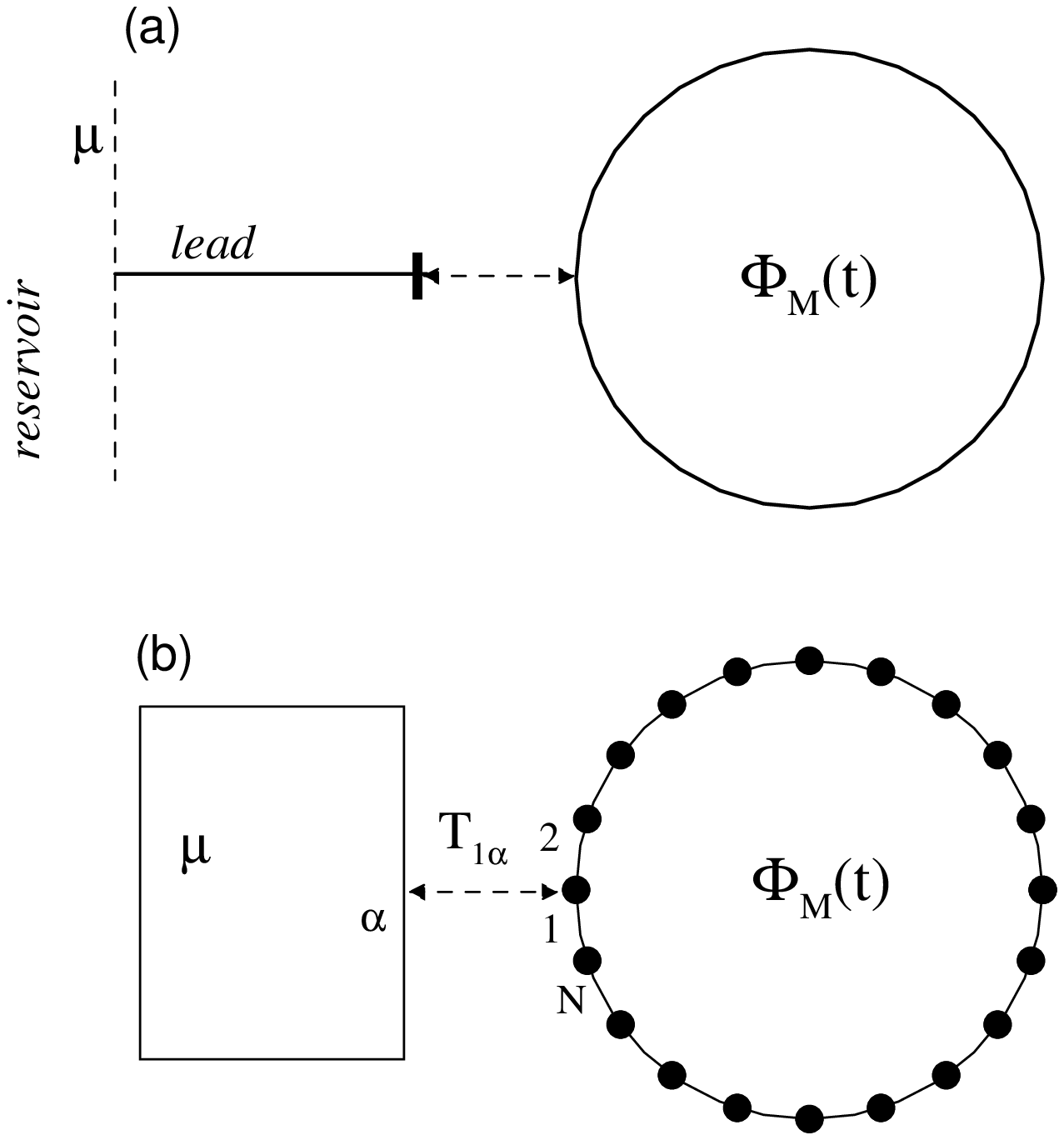 }
\caption{(a) Scheme of the experimental setup to study the
charge transport in 
a metallic ring threaded by a time-dependent magnetic field. 
(b) Theoretical model for the device in (a).}
\label{fig1a}
\end{figure}

\begin{figure}
\epsfxsize=3.5in
\epsffile{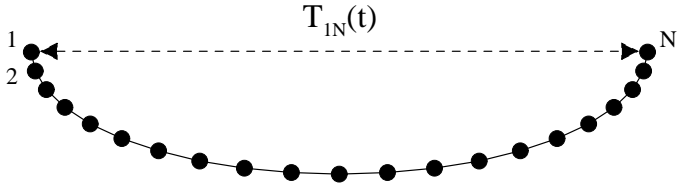}
\caption{Scheme of the physical processes involved 
in the Dyson equation for the retarded Green function (\ref{e8}).
}
\label{fig2}
\end{figure}

\begin{figure}
\epsfxsize=3.5in
\epsffile{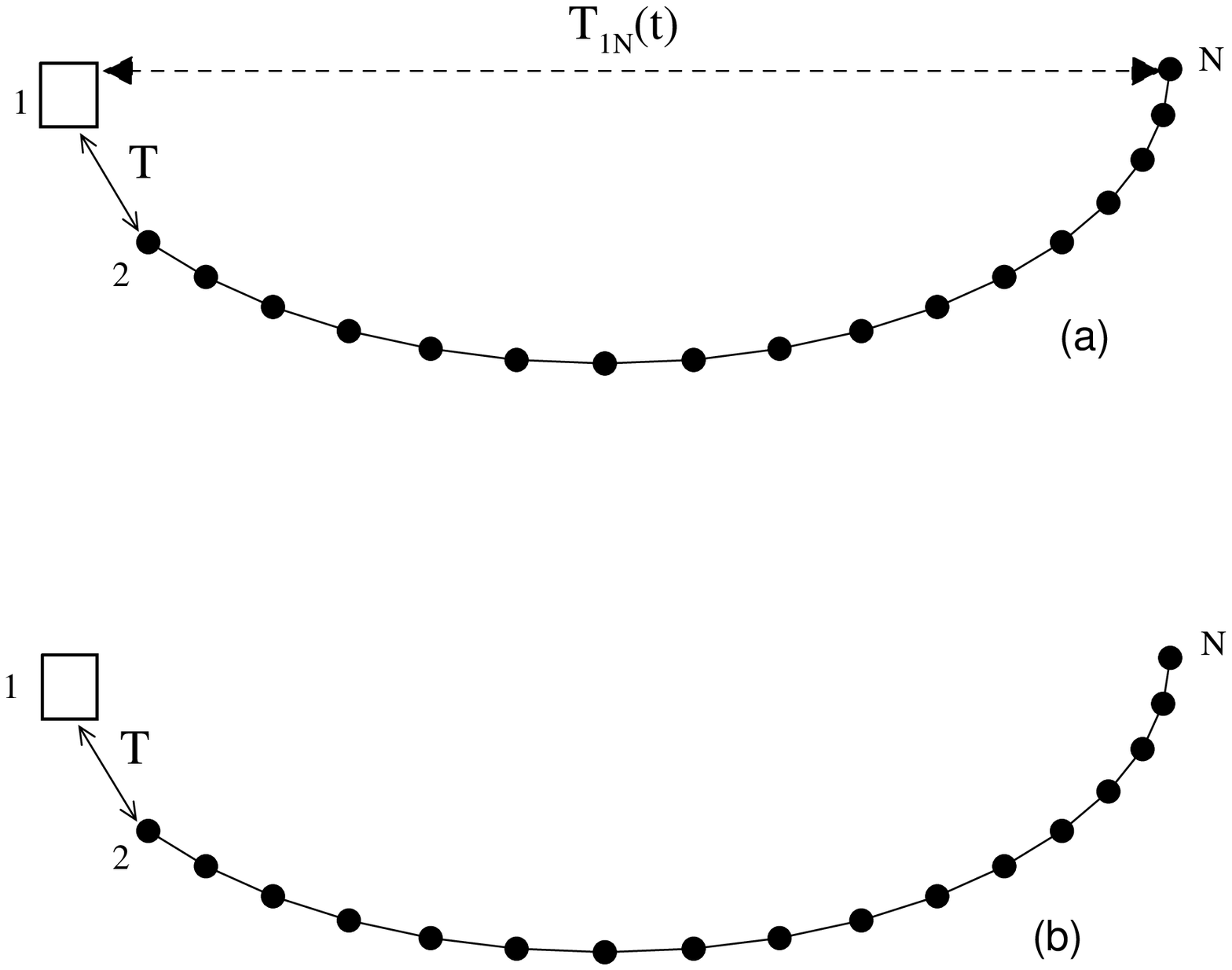}
\caption{Scheme of the physical processes involved 
in the Dyson equation for the retarded Green function.
Figs.
(a) and (b) correspond to Eqs. (\ref{ec5}) and (\ref{ec6}), respectively.
}
\label{fig3}
\end{figure}

\begin{figure}
\epsfxsize=3.5in
\epsffile{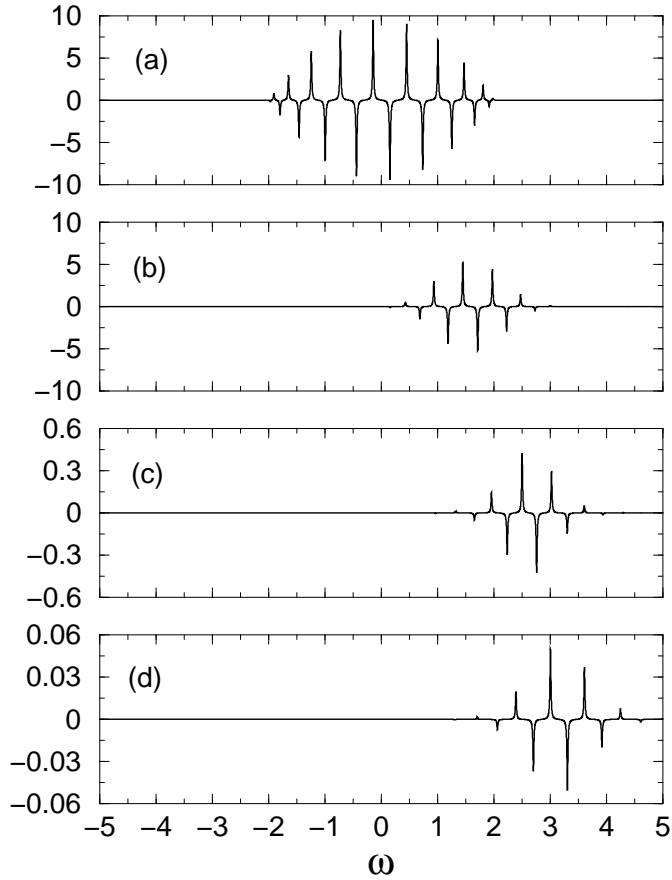}
\caption{Imaginary part of the Green function $g^0_{1N}(\omega)$
for a system of $N=20$ sites and magnetic fluxes varying at rates
$\hbar \Phi/\Phi_0=0, 0.75W, 1.25W, 1.5W$ for (a), (b), (c) and (d), 
respectively. 
}
\label{fig4}
\end{figure}

\begin{figure}
\epsfxsize=3.5in
\epsffile{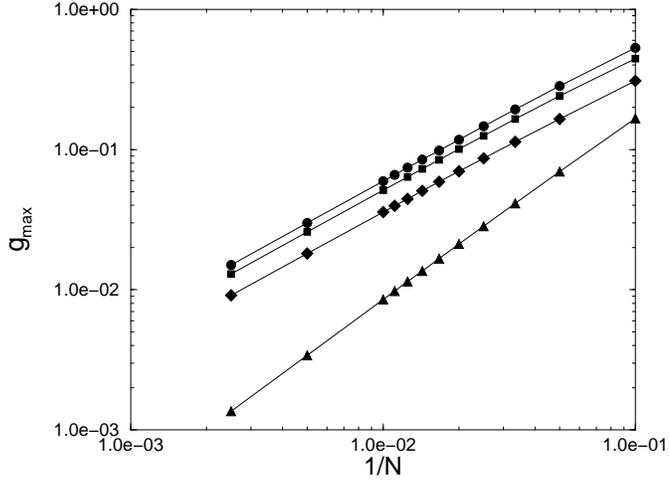}
\caption{$g_{max}$ as a function of $1/N$. Circles, squares, diamonds
 and triangles 
correspond to $\hbar \Phi/\Phi_0 = 0.25W, 0.33W, 0.5W, W$, respectively. 
}
\label{fig5}
\end{figure}

\begin{figure}
\epsfxsize=3.5in
\epsffile{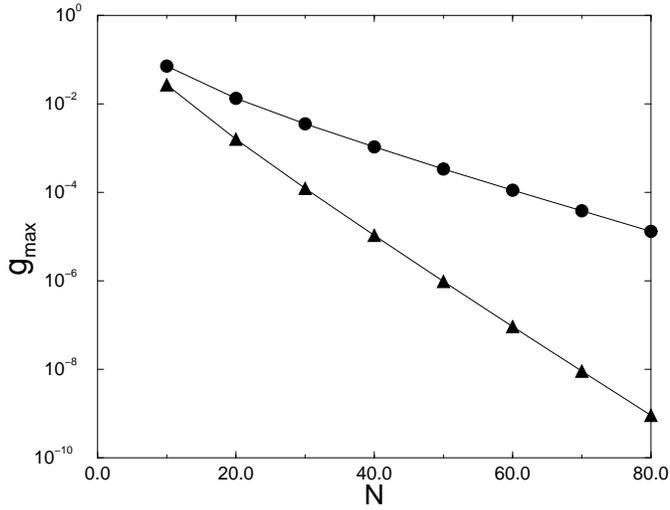}
\caption{$g_{max}$ as a function of $N$. Circles and  squares
correspond to $\hbar \Phi/\Phi_0 = 1.25W, 1.5W$, respectively. 
}
\label{fig6}
\end{figure}

\begin{figure}
\epsfxsize=3.5in
\epsffile{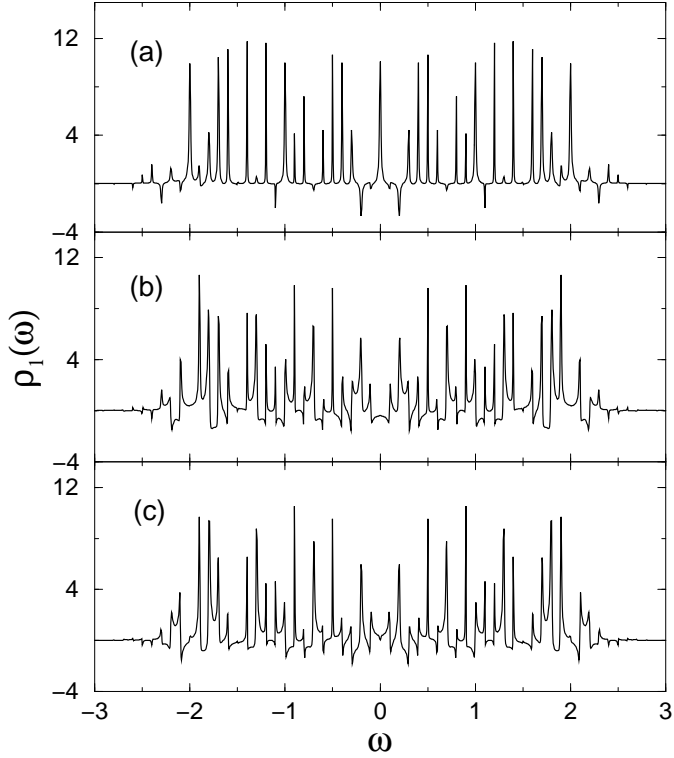}
\caption{ The generalized time-dependent density of states 
for a chain of $N=20$ sites with $\hbar \Phi/\Phi_0=0.5W$ 
$\sigma=0.025W$ at different times $t=0,1,2$ 
(a, b and c, respectively). 
}
\label{fig7}
\end{figure}

\begin{figure}
\epsfxsize=3.5in
\epsffile{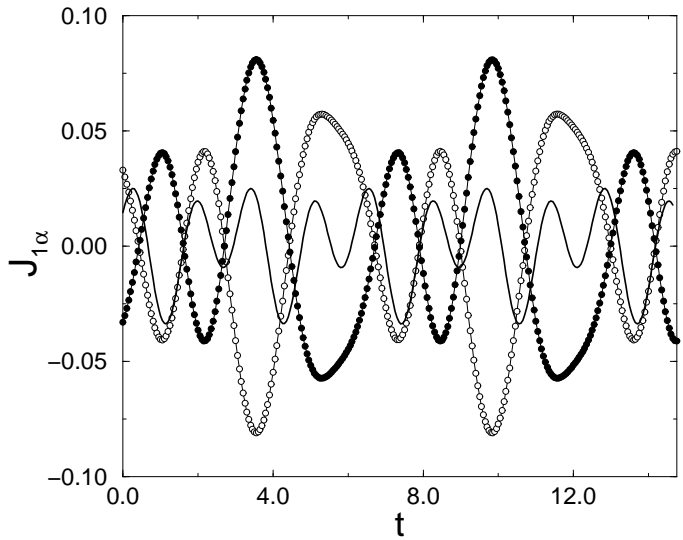}
\caption{ The current $J_{1 \alpha} (t)$
between the ring and the reservoir as a function of time
for a ring with $L=20$ sites and a self-energy $\sigma=0.2W$.
Circles correspond to
$\hbar \Phi/\Phi_0= 0.25W$ for a chemical potential of the reservoir
$\mu=0.25W$ (solid circles) and $\mu=-0.25W$ (open circles). The solid line 
corresponds to $\hbar \Phi/\Phi_0 =0.5W$ with $\mu =-0.25W$.}
\label{fig8}
\end{figure}

\begin{figure}
\epsfxsize=3.5in
\epsffile{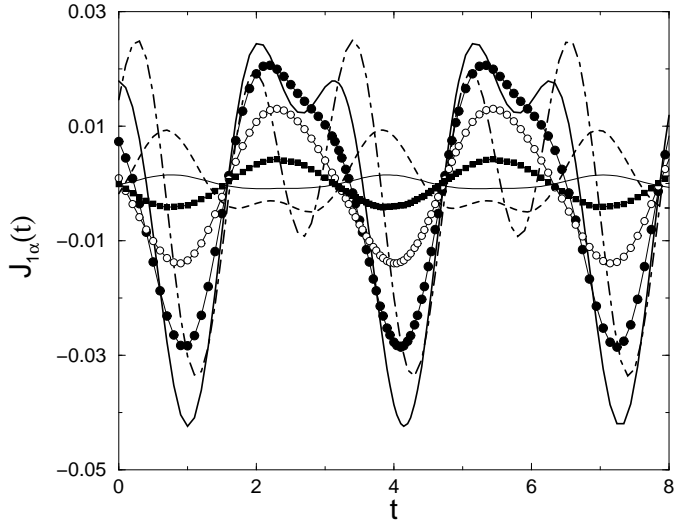}
\caption{ The current $J_{1 \alpha} (t)$
between the ring and the reservoir as a function of time
for a ring with $L=20$ sites. The  chemical potential  
of the reservoir is $\mu=-0.25W$ and  $\hbar \Phi/\Phi_0= 0.5W$.
The thin solid,
dashed, dot-dashed and thick solid lines, filled circles, opaque circles 
and filled squares 
correspond to $\sigma=0.0025W, 0.025W, 0.2W, 0.5W, W, 2.5W, 12.5W $, 
respectively.}
\label{fig9}
\end{figure}

\begin{figure}
\epsfxsize=3.5in
\epsffile{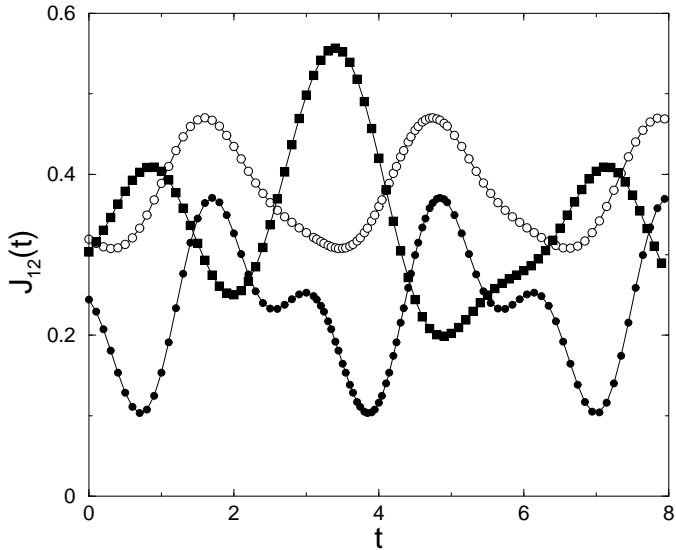}
\caption{ The current $J_{1 2} (t)$ along a ring
with $L=20$ sites and $\sigma=0.2W$. 
Circles correspond to  $\hbar \Phi/\Phi_0= 0.5W$
and a chemical potential $\mu=0$ and $\mu=-0.25W$ for opaque and
filled symbols respectively. Squares correspond to 
$\hbar \Phi/\Phi_0= 0.25W$ and $\mu=-0.25W$.}
\label{fig10}
\end{figure}

\begin{figure}
\epsfxsize=3.5in
\epsffile{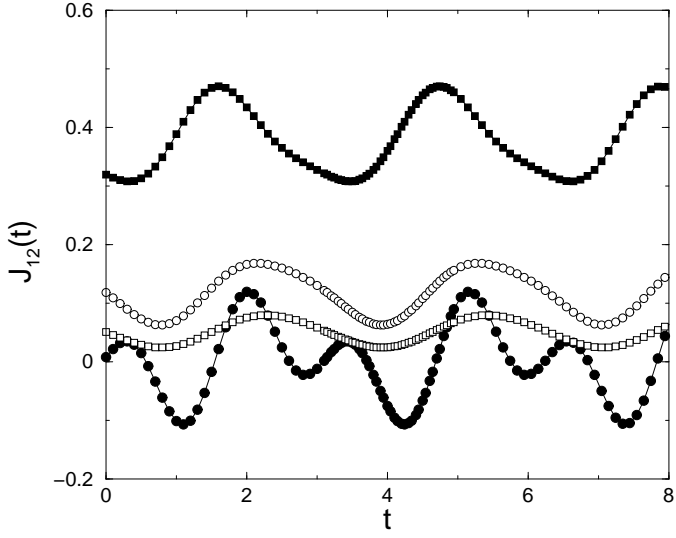}
\caption{ The current $J_{1 2} (t)$ along a ring
with $L=20$ sites and  $\hbar \Phi/\Phi_0= 0.5W$
and $\mu=-0.25W$. Filled circles, filled squares,
open circles and open squares correspond to 
$\sigma=0.0025W,0.2W,W,2.5W$, respectively.}
\label{fig11}
\end{figure}

\begin{figure}
\epsfxsize=3.5in
\epsffile{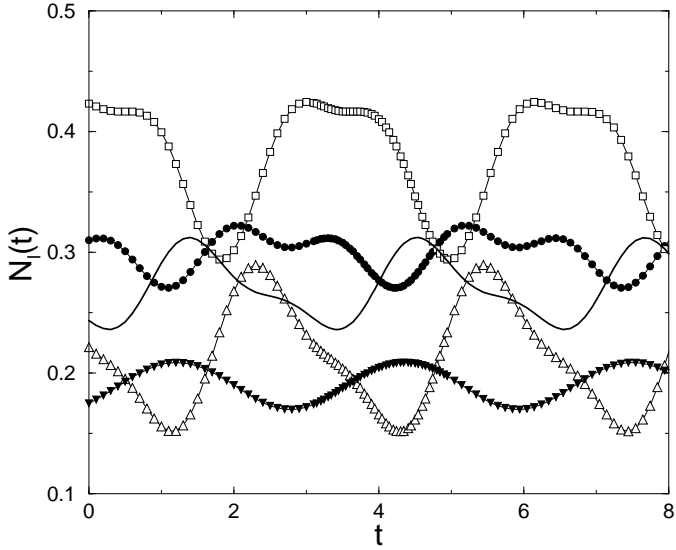}
\caption{ The occupation number $N_{l} (t)$ along a ring
with $L=20$ sites and  $\hbar \Phi/\Phi_0= 0.5W$
and $\mu=-0.25W$ and $\sigma=0.2W$. 
Circles, squares, up triangles, down triangles
and solid line correspond to $l=1,4,12,15,20$,  respectively.}
\label{fig12}
\end{figure}

\begin{figure}
\epsfxsize=3.5in
\epsffile{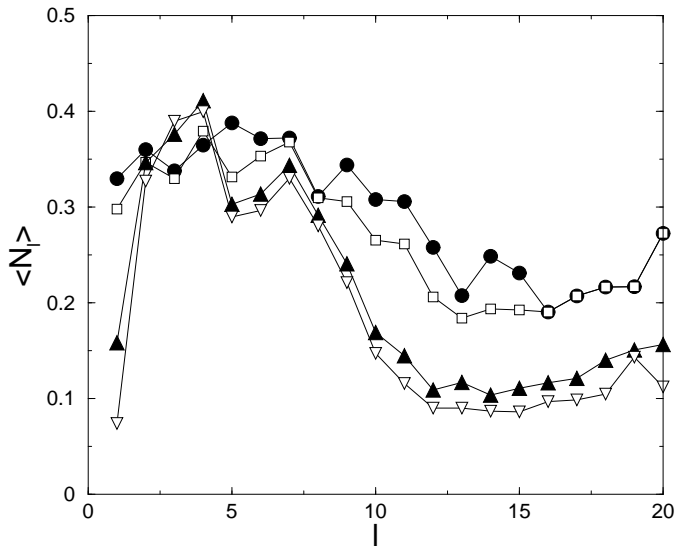}
\caption{ The mean occupation number $\langle N_{l} \rangle $ along a ring
with $L=20$ sites and  $\hbar \Phi/\Phi_0= 0.5W$
and $\mu=-0.25W$. 
Circles, squares, up triangles, down triangles
and solid line correspond to 
$\sigma=0.0025W,0.2W,W,2.5W$, respectively.}
\label{fig13}
\end{figure}


\begin{thebibliography}{99}
\bibitem{four}The analytical expresion
for the Fourier transform is 
$g^R_k(t,t^{\prime})=- i\Theta(t-t^{\prime}) \int_t^{t^{\prime}} ds 
\exp[-i \epsilon_k(s)]$,  
where $\epsilon_k(s)=-2T\cos(k+\phi s)$, with $k=2 n \pi/N, n=0,\ldots,N-1$.
\end{thebibliography}
\end{document}